\documentclass[preprint2]{aastex}

\newcommand{\myemail}{evenson@udel.edu}
\slugcomment{To Appear in Astrophysical Journal Letters}

\shorttitle{13 December 2006 Solar Particle Spectrum}
\shortauthors{Abbasi et al.}

\begin{document}


\title{Solar Energetic Particle Spectrum on 13 December 2006 Determined by IceTop}


\author{
R.~Abbasi,\altaffilmark{20} M.~Ackermann,\altaffilmark{32}
J.~Adams,\altaffilmark{11} M.~Ahlers,\altaffilmark{24}
J.~Ahrens,\altaffilmark{21} K.~Andeen,\altaffilmark{20}
J.~Auffenberg,\altaffilmark{31} X.~Bai,\altaffilmark{23}
M.~Baker,\altaffilmark{20} B.~Baret,\altaffilmark{9}
% ...
}

% ...
\altaffiltext{9}{Vrije U. Brussel, Dienst ELEM, B-1050 Brussels}

% ...
\altaffiltext{11}{Physics \& Astronomy, U. Canterbury,
Christchurch NZ}

% ...
\altaffiltext{20}{Physics, U. Wisconsin, Madison, WI 53706}

\altaffiltext{21}{Institute of Physics, U. Mainz, D-55099 Mainz}

% ...
\altaffiltext{23}{Physics \& Astronomy \& BRI, U. Delaware,
Newark, DE 19716}

\altaffiltext{24}{Physics, U. Oxford, Oxford OX1 3NP}

% ...
\altaffiltext{31}{Physics, U. Wuppertal,   D-42119 Wuppertal}

\altaffiltext{32}{DESY, D-15735 Zeuthen}

% ...




\begin{abstract}
On 13 December 2006 the IceTop air shower array at the South Pole
detected a major solar particle event. By numerically simulating
the response of the IceTop tanks, which are thick Cherenkov
detectors with multiple thresholds deployed at high altitude with
no geomagnetic cut-off, we determined the particle energy spectrum
in the energy range 0.6 to 7.6 GeV. This is the first such
spectral measurement using a single instrument with a well defined
viewing direction. We compare the IceTop spectrum and its time
evolution with previously published results and outline plans for
improved resolution of future solar particle spectra.
\end{abstract}

\keywords{Sun: flares --- Sun: particle emission}

\section{Introduction}
The IceTop air shower array now under construction at the South
Pole as the surface component of the IceCube neutrino telescope
\citep{Achterberg:2006} detected an unusual near-solar-minimum
Ground Level Enhancement (GLE) after a  solar flare on 13 December
2006. Beginning at 0220 UT, the 4B class flare occurred at solar
coordinates S06 W24, accompanied by strong (X3.4) X-ray emission
and type II and IV radio bursts. The LASCO coronagraph on the SOHO
spacecraft observed a halo CME launch from the Sun at $\sim$ 0225
UT with speed estimated to be $\sim$ 1770 km/s.  We have begun
\citep{Bieber:2007} a comprehensive analysis of the propagation of
solar energetic particles in this event. However the focus of this
Letter is the new and unique ability of IceTop to derive the
energy spectrum of these particles in the multi-GeV regime from a
single detector with a well defined viewing direction.

When completed, IceTop will have approximately 500 square meters
of ice Cherenkov collecting area arranged in an array of 80
stations on a 125 m triangular grid to detect air showers from one
PeV to one EeV. Each station consists of two, two meter diameter
tanks filled with ice to a depth of 90 cm. Tanks are instrumented
with two Digital Optical Modules (DOM) operated at different gain
settings to provide appropriate dynamic range to cover both large
and small air showers. Each DOM contains a 10 inch photomultiplier
and an advanced readout system capable of digitizing the full
waveform. For historical reasons, the two discriminator counting
rates recorded in each DOM are termed SPE (Single Photo Electron),
and MPE (Multi Photo Electron). In the present analysis the SPE
threshold corresponds approximately to 20 photoelectrons (PE), and
the MPE threshold to 100 PE.

Due to the high altitude (2835m) and the nearly zero geomagnetic
cutoff at the South Pole, secondary particle spectra at the
detector retain a significant amount of information on the spectra
of the primary particles. In a thin, ionization detector these
secondary particles either would not interact, or would produce
virtually indistinguishable signals. This is not the case in the
thick IceTop detector, where a traversing muon produces 130 PE and
the typical electron only 15 PE. Signal amplitude therefore
carries information about the composition and spectra of the
incident particles, albeit integrated over broad regions of the
spectrum. In particular, differences in counting rates of
discriminators at different thresholds allow us to infer the
particle spectrum incident at the top of the atmosphere.

\begin{figure}
\epsscale{1}\plotone{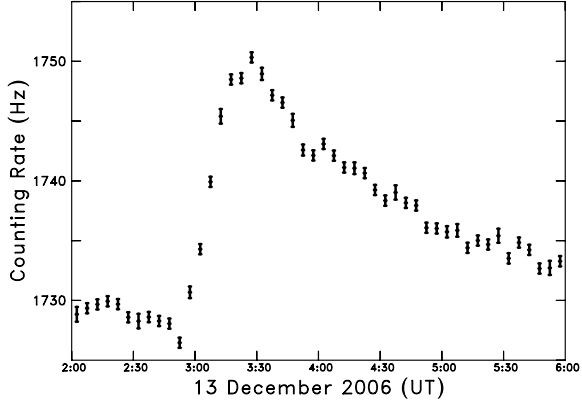}
\caption{Average Single Photo Electron (SPE) discriminator
counting rate in IceTop prior to and during the solar particle
event of 13 December 2006. Data are averaged over five minute
intervals. \label{fig1}}
\end{figure}

\begin{figure}
\epsscale{1}\plotone{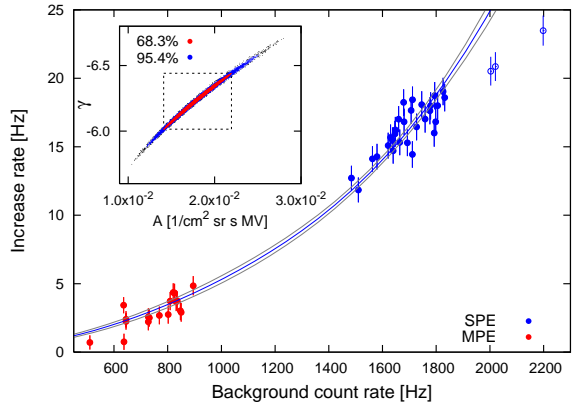}
\caption{Increase of individual IceTop discriminator counting
rates averaged from 0320UT to 0410UT over that during a reference
interval (0115UT to 0255UT) plotted against counting rate during
the reference interval. Bars indicate statistical errors. Blue
line is the best fit to a power law (in rigidity) spectrum, $A
P^{\gamma}$. Gray lines give the one sigma error band on the
spectrum as discussed in the text. Three discriminators (open
symbols) showed anomalous behavior and were excluded as described
in the text. The insert illustrates correlation of error estimates
for the fit parameters ($A, \gamma)$. \label{fig2}}
\end{figure}

\begin{figure}
\epsscale{1}\plotone{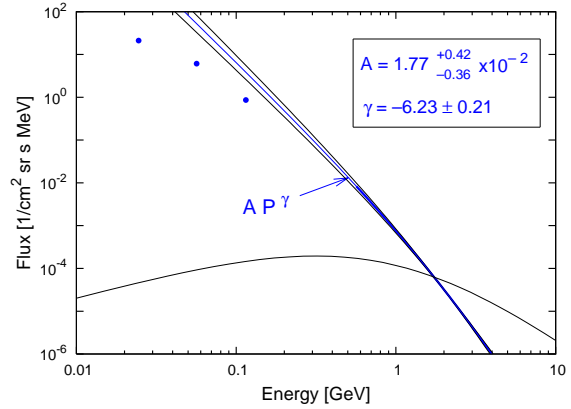}
\caption{IceTop proton spectrum from the fit in Figure~2 (heavy
blue line with one sigma error band). The black line is the
assumed background cosmic ray proton spectrum and the points are
the maximum proton fluxes from GOES spacecraft data. \label{fig3}}
\end{figure}

\begin{figure}
\epsscale{1}\plotone{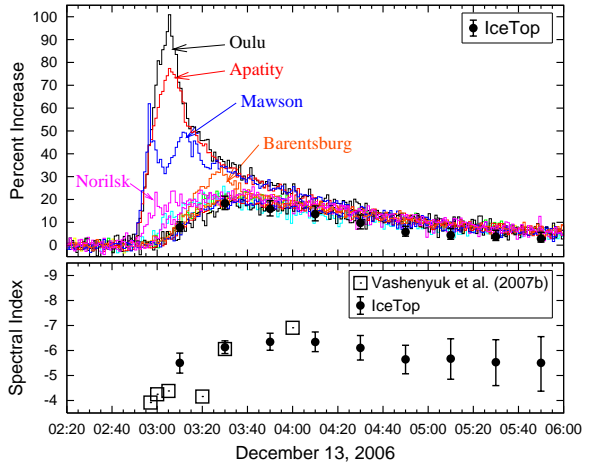}
\caption{(Top) Calculated increase in a sea level neutron monitor
based on the IceTop spectrum (heavy black circles) compared with
the counting rate of several near sea level neutron monitors.
Traces for Oulu, Apatity, Mawson, Norilsk and Barentsburg are
labelled while those for Cape Schmidt, Fort Smith, Inuvik, Nain,
Peawanuck, Tixie Bay, and Thule are not. (Bottom) Spectral index
from IceTop compared to that of \cite{Vashenyuk:2007ICRC}
\label{fig4}}
\end{figure}

\section{Observations}

On 13 December 2006, IceTop was returning useful data from 32 SPE
and 15 MPE discriminators operating in 16 tanks. Figure~1 shows
the average SPE discriminator counting rate as a function of time
with a pronounced increase due to the arrival of solar particles.
In Figure~2 we focus on the time interval 0320 to 0410 where the
spectral shape (see below) is essentially constant. We plot the
additional counting rate (due to the solar particles) for
individual discriminators as a function of counting rate prior to
the onset of the solar particle fluxes. IceTop was in test mode
with all discriminators set at nominal, uncalibrated thresholds.
Fortuitously this produced a significant gap between MPE and SPE
count rates, as well as a modest range of true thresholds within
each group. To derive an energy spectrum from these data we had to
deal simultaneously with the unknown element composition of the
particles (primarily the solar proton/alpha ratio) and the lack of
a detector calibration.

Yield functions, with units area-solid-angle, describe the
relation between particle flux at the top of the atmosphere and
the occurrence rate of a specified signal. For IceTop they depend
on the arrival direction, rigidity ($P$) and mass of the primary
nucleus and on the discriminator setting which determines the
light level required to count a particle. At high latitudes such
as South Pole defocusing in the geomagnetic field produces an
isotropic flux at the top of the atmosphere even if the flux
outside the magnetosphere is highly anisotropic. At low energy
(when the probability of a particle or its progeny to reach the
surface is small), the yield function is smaller than the physical
area-solid-angle of the tank, but at high energy (when a shower
can give rise to a signal even if the trajectory of the primary
passes outside the tank) it is larger.

By definition, the convolution of a yield function $S_{\rm pe}(P)$
with a particle spectrum $J(P)$ gives the counting rate above the
corresponding {\it pe} threshold. The product of the yield
function and the spectrum is termed the response function. Thus,
at time $t$ during the solar flare the counting rate of the {\it
i}th discriminator, with threshold pe$(i)$, is
\begin{equation}
N_i(t)\,=\, \sum_k \int\,S^k_{{\rm pe}(i)}(P)\,\{J_k(P)\,+\,\Delta
J_k(P,t)\}{\rm d}P, \end{equation} where $J(P)$ is the steady
state cosmic-ray spectrum, $\Delta J(P,t)$ is the additional flux
of particles at time t during the event, and the summation is over
particle species. Note that by interchanging summation and
integration the concept of a single response function representing
a composite spectrum is well defined.

Using FLUKA \citep{Fasso:1993}, and measured cosmic ray
composition and spectra appropriate to solar minimum, we generated
galactic cosmic ray (GCR) response functions for several different
thresholds, which could be interpolated to produce response
functions for arbitrary thresholds. Lacking a calibration, we
integrated the response functions to predict the background
counting rate for each possible threshold, then assigned to each
discriminator the response function that exactly predicted the
observed counting rates during the reference interval 0115UT to
0255UT.

Most derivations of a solar spectrum (e.g.
\cite{Bombardieri:2006}) assume the particles to be all protons
and construct a proton yield function from GCR response functions
(e.g. \cite{Lockwood:1999}). Composition at these energies has
never actually been measured, so we make the much simpler
assumption that the composition is galactic, constructing yield
functions by dividing the interpolated GCR response functions by a
modified force field proton spectrum (see Figure~3;
\cite{Caballero:2004,Clem:2004}). If solar particles are proton
rich compared to galactic, these yield functions produce a lower
limit on the solar proton spectrum. We estimate that for an all
proton composition the true intensity would be $\sim 1.2$ times
this limit.

With this set of yield functions we minimized $\chi^2$ for a power
law (in rigidity) spectrum. The actual fit is shown in Figure~2.
The minimum was 1.5 per degree of freedom, confirming the visual
impression that some of the deviation from the fit is due to
inherent differences among the detectors.  Results from a Monte
Carlo simulation are presented in an insert, with the parameter
pair resulting from each realization plotted as a point. The
68.3\% of the realizations producing the lowest $\chi^2$ are
indicated in red; we take this to define the one sigma error range
of the parameters. Due to the high correlation we do not discuss
the errors on the parameters separately. The error band shown is
the {\it envelope} of all possibilities within the 68.3\% contour.
Three discriminators (open symbols in Figure~2) were excluded from
the fit primarily because their background counting rates stand so
far outside the cluster of discriminators operating under the same
nominal conditions. Correlation of their background count rate
with barometer reading over the three day interval surrounding the
flare event was also anomalous compared to that of the other
discriminators.

The derived IceTop spectrum is shown in Figure~3 by the blue
curve, with the heavier line denoting the energy range that
contributes substantially (tenth to ninetieth percentile) to the
fit. The parameter ranges quoted correspond to the dashed
rectangle in Figure~2. The overshoot of the spectral extrapolation
compared to low energy proton fluxes from the GOES spacecraft is
rather typical, and is generally interpreted as a steepening of
the spectrum over the intervening energy range. Note that IceTop
is able to derive the spectrum of a small increase over a large
background. An increase of this magnitude would not be
statistically resolved with a detector of a size practical for
flight on a spacecraft or balloon.

\section{Interpretation}

Repeating this analysis on 20 minute subsets of the data, and
using a published response function \citep{Moraal:1989} we
calculated (Figure~4) the expected count rate increase for a sea
level neutron monitor. The error bars were determined, as in
Figure~2, as the range of allowed values from parameter pairs
within a 68.3\% contour. If, instead, we were to compute the
standard deviation over all realizations in the Monte Carlo
simulation, the error bars would be a factor $\sim 0.65$ smaller.
For comparison we show one minute averages of the counting rates
of several near sea level neutron monitors. Particles first
arrived from the sunward direction, observed best by Oulu, Mawson,
and Apatity in a very tight beam focused by the interplanetary
magnetic field \citep{Bieber:2007}. The initial rise was seen by
Mawson, then the beam moved over to Oulu and Apatity, before
briefly switching back toward Mawson.  By the time Barentsburg saw
the beam, nearly isotropic scattered particles were beginning to
dominate the flux. The viewing direction of IceTop was similar to
that of monitors that primarily observed backscattered particles.
We consider the agreement of our calculation with the observations
from these monitors to be rather good, given that we arguably have
no free parameters.

Traditional methods for determining energy spectra rely on
observations from pairs or groups of stations with different
geomagnetic cutoff and station altitude
\citep{Lockwood:2002,Ryan:2005,Bombardieri:2006} which typically
have strongly energy dependent viewing directions. Such an
analysis has been reported for the 13 December 2006 event by
\cite{Vashenyuk:2007ASR, Vashenyuk:2007ICRC} who employ the world
network of neutron monitors to achieve a range of viewing
directions and geomagnetic cutoffs under the common assumption
that the event can be modelled by a function separable in energy
and anisotropy. In the lower panel of Figure~4 we compare the
spectral index derived from both analyses. Late in the event, when
the fluxes are more isotropic, the agreement is excellent. Early
on, \cite{Vashenyuk:2007ICRC} derived a significantly harder
overall spectrum than ours, and also reported a pronounced
softening of the spectrum with time. In contrast, our analysis
yielded a nearly constant spectral index. We believe that the
discrepancy results from the way \cite{Vashenyuk:2007ICRC}
parameterized the anisotropy. As a result, their fit confuses
anisotropy evolution with spectral evolution. We are confident
that we eventually will converge on a common understanding when
the precise spectrum derived from IceTop is properly included as a
constraint on the fit. When definitive results from the PAMELA
spacecraft instrument are available they should also contribute
greatly to a comprehensive analysis. It is also clear that future
results from IceTop will be greatly enhanced by the neutron
monitor network, which will continue to be the primary source of
information on anisotropy.

\section{Future Plans for IceTop}

Encouraged that such a straightforward analysis of IceTop data
yields a useful picture of the time dependent spectrum of the
solar particle event on 13 December 2006, we are working to better
understand the instrumentation to reduce systematic uncertainties.
We are also reconfiguring IceTop to increase statistical
precision. DOMs are being reprogrammed to collect and transmit
histograms, with ten second resolution, of the integrated charge
signal for all events that trigger the MPE discriminators. These
will then be set to trigger at a rate of approximately 2000 Hz, a
limit determined by the tolerable dead time of the system for air
shower studies. Each of the (eventual) 160 high gain DOMs will
thus return a spectrum that is statistically equivalent to the
spectrum returned by the entire ensemble of DOMs employing the
present data collection method. SPE discriminators will be set at
a variety of thresholds to populate the regime at rates higher
than 2000 Hz, probably up to about 10,000 Hz. DOMs are capable of
accumulating histograms at these higher rates, but the dead time
would be unacceptable for normal operation. Overall we should be
able to achieve a sensitivity two orders of magnitude greater than
current neutron monitors for event detection.

These changes will dramatically improve the energy resolution of
IceTop for determining solar flare spectra. In Figure 2, there is
some spread within the two clusters of points but not enough to go
much beyond a two parameter fit. With properly spaced coverage and
extensions out to 10,000 Hz, we will be able to measure spectral
curvatures or cutoffs. We are indeed looking forward to the coming
solar maximum.

\acknowledgments

We acknowledge the support from the following agencies: National
Science Foundation Office of Polar Programs; National Science
Foundation Physics Division; University of Wisconsin Alumni
Research Foundation; Department of Energy; National Energy
Research Scientific Computing Center (supported by the Office of
Energy Research of the Department of Energy); NSF-supported
TeraGrid system at the San Diego Supercomputer Center (SDSC);
National Center for Supercomputing Applications (NCSA); Swedish
Research Council; Swedish Polar Research Secretariat; Knut and
Alice Wallenberg Foundation; German Ministry for Education and
Research; Deutsche Forschungsgemeinschaft (DFG); Fund for
Scientific Research (FNRS-FWO); Flanders Institute to encourage
scientific and technological research in industry (IWT); Belgian
Federal Science Policy Office; Netherlands Organisation for
Scientific Research (NWO); M. Ribordy:  SNF (Switzerland); A.
Kappes:  EU Marie Curie OIF Program; T. Kuwabara and J. W. Bieber:
NASA grants NNX07AH73G and NNX08AQ18G. We thank our colleagues at
IZMIRAN, Polar Geophysical Institute (Russia), and Australian
Antarctic Division for furnishing neutron monitor data.


\clearpage

\end{document}